 \documentstyle[twocolumn,prl,aps]{revtex}

\begin{document}
\draft
\preprint{}


\title{Spin dynamics of dimerized Heisenberg chains}
\author{Andreas Fledderjohann$^1$ and Claudius Gros$^2$
       }
\address{$^1$ Physics Department, University of Wuppertal,
         42097 Wuppertal, Germany
        }
\address{$^2$ Institut f\"ur Physik, Universit\"at Dortmund,
         44221 Dortmund, Germany
        }
\date{\today}
\maketitle
\begin{abstract}
We study numerically the dimerized Heisenberg
model with frustration appropriate for the
quasi-1D spin-Peierls compound CuGeO$_3$.
We present evidence for a bound state in
the dynamical structure factor for any finite
dimerization $\delta$ and estimate the respective
spectral weight. For the homogeneous case
($\delta=0$) we show that the spin-wave velocity
$v_s$
is renormalized by the n.n.n. frustration term
$\alpha$ as
$v_s=\pi/2\,J(1-b\alpha)$ with $b\approx1.12$
\end{abstract}

\bigskip

Quantum 1D spin systems may show a variety
of instabilities. Of particular interest is the
spin-Peierls (SP) phase due to residual
magnetoelastic couplings \cite{Beni_Pincus},
which leads to the opening of a gap in the
spin excitation spectrum.
The discovery \cite{Hase} of the spin-Peierls 
transition below $T_{SP}=14$ K 
in an inorganic compound, CuGeO$_3$, 
has attracted widespread attention. 

The spin-dynamics of
CuGeO$_3$ is being studied intensively
\cite {Nishi,Martin,Lorenzo}. Above 
the spin-Peierls transition the 
one-magnon excitation spectrum,
as measured by neutron scattering
\cite{Kakurai}, forms a broad continuum.
The form of the continuum is in good
agreement with the continuum expected
for one-dimensional antiferromagnetic Heisenberg chains
\cite{Mueller}; it is also seen in
the magnetic Raman spectrum \cite{Raman,Muthu}.
The physics behind the continuum in the
one-magnon excitation spectrum lies
on the fact that the elementary excitations
of the one-dimensional Heisenberg chain
are spinons (also called solitons or
Bloch walls), each magnon being made
up of two solitons. The resulting
``two-spinon'' continuum has been seen
experimentally also in other quasi-1D
antiferromagnets \cite{KCuF3}.

The exact form of the magnetic
excitation spectrum in the dimerized
state below $T_{SP}$ is still being
investigated. A two-spinon continuum
is observed \cite{Lorenzo,Kakurai} with
large spectral weight at the lower
edge. The dispersion of the lower edge
has been used to extract
the spin-wave velocity \cite{Nishi}.
Muthukumar {\it et al.} have pointed out
recently \cite{Muthu} that the experimental
magnetic Raman intensity indicates a
well defined magnon mode in the spin-Peierls
state of CuGeO$_3$ and that the origin of 
this magnon could not be determined on  the
basis of the Raman scattering data
obtained numerically for a single chain.

The existence of a well defined magnon mode,
{i.e.} a two-spinon bound state which splits of
the continuum, has
also been addressed recently by
Uhrig and Schulz \cite{Uhrig} within an
RPA approach. A recent neutron scattering experiment \cite{Ain},
has been interpreted as indicative of
such a bound-state.

In this context it is an important question to address
whether other techniques can shed light on
the magnetic excitation spectrum of the
dimerized Heisenberg model. Here we present
data for the dynamical structure factor
$S(k,\omega)$ obtained by applying the
recursion method \cite{rec} to calculate
the excitation spectrum of the hamiltonian
given below. This approximate method in
particular gives very accurate results
for the low-lying excitaions. We find
evidence for a bound-state for any dimerization
$\delta>0$, independently of the amount of
frustration $\alpha$ present. The finite-size corrections
of the data is, for certain k-values, small
enough to determine accurately the position of
the bound state and its spectral weight.
We furthermore determine the renormalization of the
spin-wave velocity in the homogeneous state as a function
of the frustration.
\medskip

The magnetic properties of CuGeO$_3$ 
can be modeled by the 1D spin-Hamiltonian
\begin{equation}
H = J\sum_i\  [ (1+\delta(-1)^i)\,{\bf S}_i\cdot{\bf S}_{i+1}
            +    \alpha       {\bf S}_i\cdot{\bf S}_{i+2}
              ],
\label{H}
\end{equation}
where $\delta$ is the dimerization parameter that 
vanishes above $T_{SP}$, its zero temperature
value has been estimated to be $\delta=0.03$
\cite{Castilla}.
The exchange integral $J>0$ and the
amplitude for the frustrating n.n.n. coupling
$\alpha$ have been estimated
to be $J\approx150$K \cite{Castilla}
and $\alpha\approx0.24-0.36$ \cite{Castilla,Riera},
respectively.

The phase diagram of $H$ in Eq.\ (\ref{H})
has been calculated using the 
density-matrix renormalization-group method \cite{Chitra}.
For $\delta=0$ and $\alpha<\alpha_c\approx0.2411$,
the ground state is gapless and renormalizes to the
Heisenberg fixed point. The rest of the phase
space is gapped.  For $\delta+2\alpha=1$
the exact ground state is known to be a valence-bond
state.

The dynamical structure factor $S(k,\omega)$ in the
ground-state $|0\rangle$ of (\ref{H}) is given by
\begin{equation}
S(k,\omega) = \sum_n \,|\langle0|S^z(-k)|n,k\rangle|^2\,
              \delta(\omega-(E_n(k)-E_0)),
\label{S}
\end{equation}
where $|n,k\rangle$ and $E_n(k)$ are the respective
eigenstates and the eigenvalues of (\ref{H})
in the subspace with total momentum $k$,
$|0\rangle$ being the overall ground-state
and $E_0$ the ground-state energy. 
The allowed values for the
momentum $k$ are $k=0,\, 2\pi/N,\dots,\, \pi$,
for a finite chain of length $N$
with periodic boundary conditions.
$S^z(k)$ is given by
\[
S^z(k) \, =\, {1\over\sqrt{N}}\sum_{x=1}^N\,\exp(ikx)\,S_x^z.
\]
We have evaluated $S(k,\omega)$ for chains with 
$N=12,\, 16,\, 20,\, 24$ using an approximate scheme
for the determination of the lowest-lying excitation
energies $E_n(k,N)-E_0(N)$ and the corresponding
transition probabilities $|\langle n|S^z(k)|0\rangle|^2$
\cite{rec}. Using a recursion algorithm a set of orthogonal
states is built starting with $S^z(k)|0\rangle$ \cite{rhi}.
Coefficients occuring in this procedure form a tridiagonal
matrix whose eigenvalues and eigenstates determine
the excitation energies and transition probabilities.
(For further details and technichal limitations of
this method see ref. \cite{rec}).

The dynamical structure factor is given, for any finite chain,
by a sum over discrete poles. The weight of the individual
poles,
\[
w_n(k)\, =\, |\langle0|S^z(-k)|n,k\rangle|^2
\]
will go to zero in the thermodynamic limit,
$N\rightarrow\infty$, whenever the respective pole
contributes to the continuum.
The sum-rule
\begin{equation}
\sum_n\, w_n(k)\, =\, S(k)
\label{sum_rule}
\end{equation}
is here valid,
where $S(k)$ is the static structure factor,
\begin{equation}
S(k)\, = \, \langle0|S^z(-k)\,S^z(k)|0\rangle.
\label{S_k}
\end{equation}
We are here interested in particular in
the lowest pole, located
at $E(k)=E_0(k)-E_0$, where $|n=0,k\rangle$ is the
lowest state with energy $E_0(k)$
in the subspace of total momentum $k$.
We want to examine the
question whether this pole is part of the continuum
in the thermodynamic limit, or whether it evolves
into a bound state, characterized by
\begin{equation}
w_0(k) \rightarrow {\rm const.},\qquad N\rightarrow\infty.
\label{w}
\end{equation}
In Fig.\ (\ref{alpha=0}) we present for $\alpha=0$
the relative weight of the lowest pole
contributing to $S(k,\omega)$,
\begin{equation}
w_0^{(rel)}(k)\,=\,w_0(k)\,/\, S(k).
\label{w^rel}
\end{equation}
The data for $w_0^{(rel)}(k)$
presented in  Fig.\ (\ref{alpha=0}) 
shows a monotonic decrease for
$\delta=0$ as a function of chain
length $N$, indicating that the lowest pole
will be part of the
two-spinon continuum in the thermodynamic
limit. This behaviour is what we expect
for the homogeneous Heisenberg chain.
For the dimerized Heisenberg chain,
$\delta=0.1$ in Fig.\ (\ref{alpha=0}),
we observe on the other hand for
all $k<\pi$ a monotonic
{\it increase} of $w_0^{(rel)}(k)$
with increasing chain length!
This behaviour is also obtained when we 
introduce a finite amount of frustration $\alpha=0.24$
into the system, as can be seen
in Fig.\ (\ref{alpha=0.24}), where we compare
the data for $\delta=0$ and $\delta=0.03$, as
apropiate for CuGeO$_3$.

We also observe in
Fig.\ (\ref{alpha=0}) and Fig.\ (\ref{alpha=0.24})
that the finite-size dependence of the
relative weight of the lowest pole is 
small enough in the dimerized state for $k\le\pi/2$, that
an estimate of the spectral weight in the thermodynamic
limit is possible for $k\le\pi/2$.  Taken
together we conclude that the data presented in
Fig.\ (\ref{alpha=0}) and Fig.\ (\ref{alpha=0.24})
clearly indicate for $\delta>0$
a finite value for $w_0(k)$ in 
the thermodynamic limit and therefore a bound state.

Another interesting feature of the data presented in
Fig.\ (\ref{alpha=0}) and Fig.\ (\ref{alpha=0.24})
is the large dependence of $w_0^{(rel)}(k)$ on
the dimerization parameter $\delta$. This 
observation is reminiscent to the large
$\delta$-dependence observed for the Raman
spectral weight in a previous study \cite{Muthu}.
It is interesting to note that this behaviour
is solely due to the $\delta$-dependence of the
matrix element $w_0(k)$ and that the
static structure factor $S(k)$,
entering Eq.\ (\ref{w^rel}) is relatively
size and $\delta$-independent for most
$k<\pi$, as can been seen form the data
presented in Fig.\ (\ref{static}).

In Fig.\ (\ref{static}) we present for
$\alpha=0.24$ the data for static structure 
factor $S(k)$ and the energy dispersion
$E(k)=E_0(k)-E_0$ for both $\delta = 0$ 
and $\delta=0.03$ and
for $N=20$ (open symbols) and $N=24$
(filled symbols). We notice that the
finite size effects are, in general, much
smaller than those for the weight
of the lowest pole presented in the
previous figures and that the finite
size effects for $\delta=0.03$ are
smaller than those for $\delta=0$.

A relative large difference for $\delta=0$ and
$\delta=0.03$ occurs, for the data presented
in Fig.\ (\ref{static}) for the 
``magnon dispersion'' $E(k)$, for
$k>\pi/2$. This behaviour is a consequence of
the coupling of the momenta $k$ and $k+\pi$ by the
dimerization. For any finite
$\delta>0$ we have therefore the symmetry
$E(\pi/2+\Delta k)=E(\pi/2-\Delta k)$.
In the homogeneous
case, $\delta=0$, one has instead that
$E(\pi/2+\Delta k)>E(\pi/2-\Delta k)$
for any finite chain and $\pi/2>\Delta k > 0$.

In Fig.\ (\ref{v_s}) we present a study of the
spin-wave velocity, $v_s$, as a function of
frustration $\alpha$ and a range of system
sizes $N=12,\, 16,\, 20,\, 24$, and $\delta=0$. Here we 
estimated $v_s$ using the formula
\[
v_s\ =\ {E(k)\over k}\,\Big|_{k=2\pi/N}.
\]
Strictly speaking we expect $v_s$ to diverge
like $\sim\Delta N/2\pi$
in the thermodyamic limit for
any $\alpha>\alpha_c\approx0.2411$ where a
gap $\Delta$ opens in the spin-wave spectrum.
This divergence will be very difficult to
observe in finite-size studies as the 
gap is exponentially small for $\alpha$
larger but close to $\alpha_c$ \cite{Chitra}.

The size-dependence of $v_s$ is evident also
for small values of $\alpha$. For $\alpha=0$
the exact values for the spin-wave velocity
is known to be $\pi/2J$ in the thermodynamic,
given by the intersect of the solid line
in Fig.\ (\ref{v_s}) with the y-axis. The finite
size corrections do, on the other hand, actual
increase slighty with increasing chain length
$N$, at least for $N\le24$ and $\alpha=0$.
We have therefore decided to try for a linear
fit of $v_s$ as a function of $\alpha$ by demanding
the fit to reproduce the Bethe-Ansatz value
for $\alpha=0$. The corresponding fit 
(solid line in Fig.\ (\ref{v_s}) is
\[
v_s\ =\ {\pi\over2} J\,(1-b\alpha),
\]
with $b\approx1.12$ \cite{note}.

\bigskip
In conclusion we have presented numerical evidence
for the ocurrence of a discrete-pole contribution
to the dynamical structure factor for dimerized
Heisenberg chains. This contribution to $S(k,\omega)$
may also be interpreted, at least in part,
as a two-spinon bound state \cite{Uhrig},
it constitutes a well-defined magnon mode.
It is interesting to note that such a bound-state
would imply a peak in the two-magnon density of
states and might therefore show up in the
Raman-spectral weight. A previous numerical
numerical study has {\it not} found such a peak
structure in the Raman spectral weight of
dimerized Heisenberg chains. This discrepancy
may be either due to strong magnon-magnon
interaction effects in 1D spin chains or
due to matrix-element effects
\cite{Muthu,Singh}.

We have furthermore presented data for
the dispersion of the magnon-mode and the
static structure factor, showing that the
finite-size effects are small for these
quantities. In addition we have estimated
the renormalization of the spin-wave 
velocity for frustrated Heisenberg chains
in the gapless region.


\acknowledgements
This work was supported by the Deutsche 
For\-schungs\-gemein\-schaft, 
the Graduierten\-kolleg ``Fest\-k\"{o}rper\-spekt\-roskopie''.

%
%
%
\begin{figure}
\caption{The relative weight $w_0^{(rel)}(k)$ of the
         lowest pole contributing to the dynamical structure factor
         $S(k,\omega)$
         for $\alpha=0$, $\delta = 0$ and $\delta=0.1$ and
         for chains with $N=16,\, 20,\, 24$ sites respectively.
         The relative weight ist given by
         $w_0^{(rel)}(k)=w_0(k)/\sum_n w_n(k)$, where
         the $w_n(k)$ are the absolute weights of the poles
         contributing to $S(k,\omega)$.
         The lines are guides to the eye.
\label{alpha=0}
             }
\end{figure}
\begin{figure}
\caption{The relative weight $w_0^{(rel)}(k)$ of the
         lowest pole contributing to the dynamical structure factor
         for $\alpha=0.24$, $\delta = 0$ and $\delta=0.1$ and
         for chains with $N=16,\, 20,\, 24$ sites respectively.
         The lines are guides to the eye.
\label{alpha=0.24}
             }
\end{figure}
\begin{figure}
\caption{The energy of the lowest excited state
         with momentum k, E(k), and the static
         structure factor, S(k), for $\alpha=0.24$
         and both $\delta=0$ and $\delta=0.03$.
         The open/filled symbols are the data for chains
         with $N=20$ and $N=24$ sites respectively.
         The lines are guides to the eye.
\label{static}
             }
\end{figure}
\begin{figure}
\caption{The spin-wave velocity, $v_s$, as a function of $\alpha$
         calculated for $\delta=0$ and
         chains with length $N=12,\, 16,\, 20,\, 24$.
         $v_s$ has been determined using the formula
         $v_s=(E(2\pi/N)-E(0))/(2\pi/N)$, where
         $E(k)$ is the ground-state energy for
         chains with total momentum $k$.
         The dahsed lines are guides to the eye, 
         the full line is a fit.
\label{v_s}
             }
\end{figure}

\begin{references}

\bibitem{Beni_Pincus} L. N. Bulaevski\u{i},
		      Sov. Phys. JETP {\bf 17}, 684 (1963);
                      G.~Beni and P. Pincus,
                      J. Chem. Phys. {\bf 57}, 3531 (1972).

\bibitem{Hase} M. Hase {\it et al.}, 
               Phys. Rev. Lett. {\bf 70}, 3651 (1993).

\bibitem{Nishi} M. Nishi, O. Fujita and J. Akimitsu,
                Phys. Rev. B {\bf 50}, 6508 (1994).

\bibitem{Martin} M.C. Martin {\it et al.},
                 Phys. Rev. B {\bf 53}, R14 713 (1996).

\bibitem{Lorenzo} J.E. Lorenzo {\it et al.},
                  {\it ``Spin dynamics in the Spin Peierls
                  compound CuGeO$_3$''}, preprint.

\bibitem{Kakurai} N. Kakurai, 
                  Bull. Am. Phys. Soc. Vol. 41, 625 (1996).

\bibitem{Mueller} See, e.g. G. M\"uller, H. Thomas, H. Beck
                  and J.C. Bonner,  
                  Phys. Rev. B {\bf 24}, 1429 (1981);
                  M. Karbach, G. M\"uller and A.H. Bougourzi,
                  Sissa preprint cond-mat/9606068.

\bibitem{Raman} H. Kuroe {\it et al.},
                Phys. Rev. B {\bf 50}, 16 468 (1994);
                P.H.M. van Loosdrecht {\it et al.},
                Phys. Rev. Lett. {\bf 76}, 311 (1996);
                Lemmens {\it et al.} preprint;
                H. Ogita {\it et al.}, preprint. 

\bibitem{Muthu} V.N. Muthukumar {\it et al.}, 
                {\it ``Frustration induced Raman scattering
                in GeCuO$_3$''}, preprint.

\bibitem{KCuF3} D.A. Tennant, R.A. Cowley, S.E. Nagler and
                A.M. Tsvelik,
                Phys. Rev. B {\bf 52}, 13 368 (1995).

\bibitem{Uhrig} G.S. Uhrig and H.J. Schulz,
                Sissa preprint cond-mat/9606001.

\bibitem{Ain} M. A\"\i n, J.E. Lorenzo, L.P. Regnault,
              G. Dhalenne, A. Revcolevschi and Th. Jolicoeur,
              preprint.

\bibitem{rec}
        A. Fledderjohann, M. Karbach, K.-H. M\"utter and P. Wielath,
                {J. Phys.: Condens. Matter} {\bf 7}, 8993 (1995)
\bibitem{rhi}
        V.S. Viswanath, S. Zhang, J. Stolze and G. M\"uller,
                {Phys. Rev.} {\bf B} 49, 9702 (1994) \\
        V.S. Viswanath and G. M\"uller,
                {\it The Recursion Method - Application to Many Body Dynamics}
                Springer Lecture Notes in Physics m 23, Springer-Verlag,
                New York (1994)


\bibitem{Castilla} G. Castilla, S. Chakravarty and V.J. Emery,
                   Phys. Rev. Lett. {\bf 75}, 1823 (1995).

\bibitem{Riera} J. Riera and A. Dobry,
                Phys. Rev. B {\bf 51}, 16 098 (1995).

\bibitem{Chitra} R. Chitra {\it et al.},
                 Phys. Rev. B {\bf 52}, 6581 (1995).

\bibitem{Singh} R.R.P. Singh, P. Prelovsek and B. S. Shastry,
                {\it ``Magnetic Raman Scattering form 1D
                Antiferromagnets''}, preprint.

\bibitem{note}
This number may be
compared with the formula
$v_s= (1+2/\pi)J[1-4\alpha/(2+\pi)]$
found within the framework of the 
solitonic mean-field theory
\protect\cite{Muthu}.

%
%
%
%
%

\end{references}
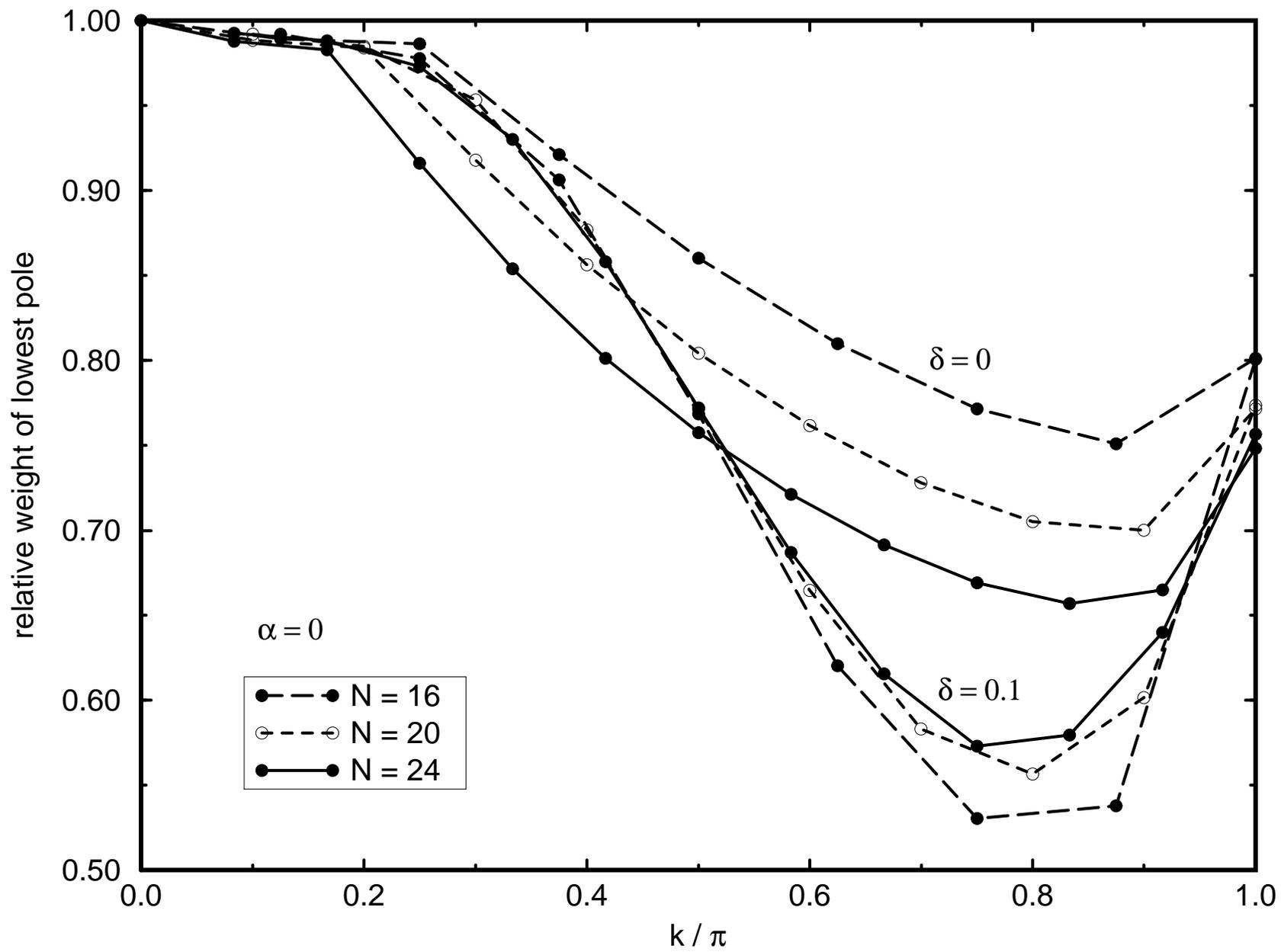
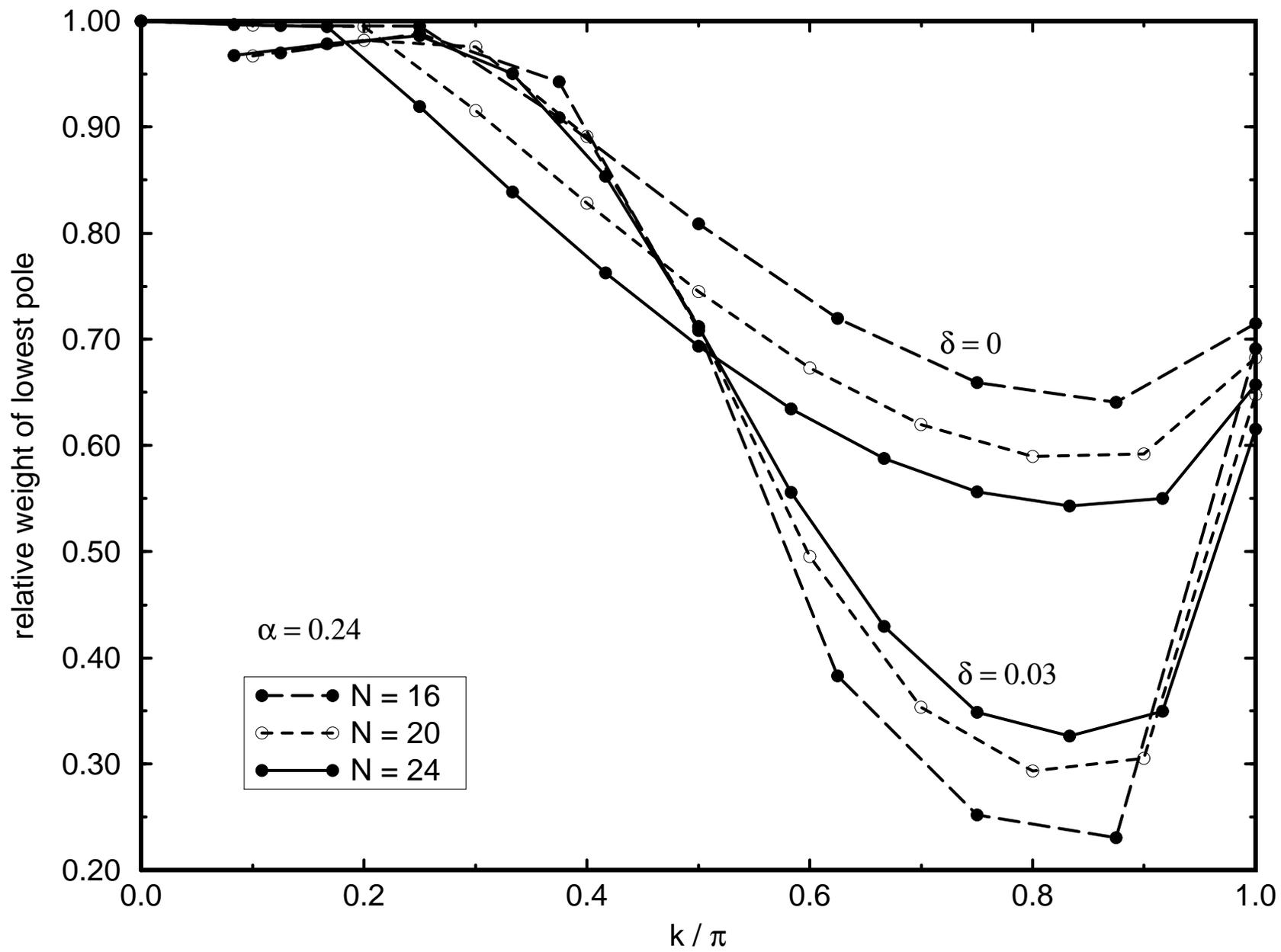
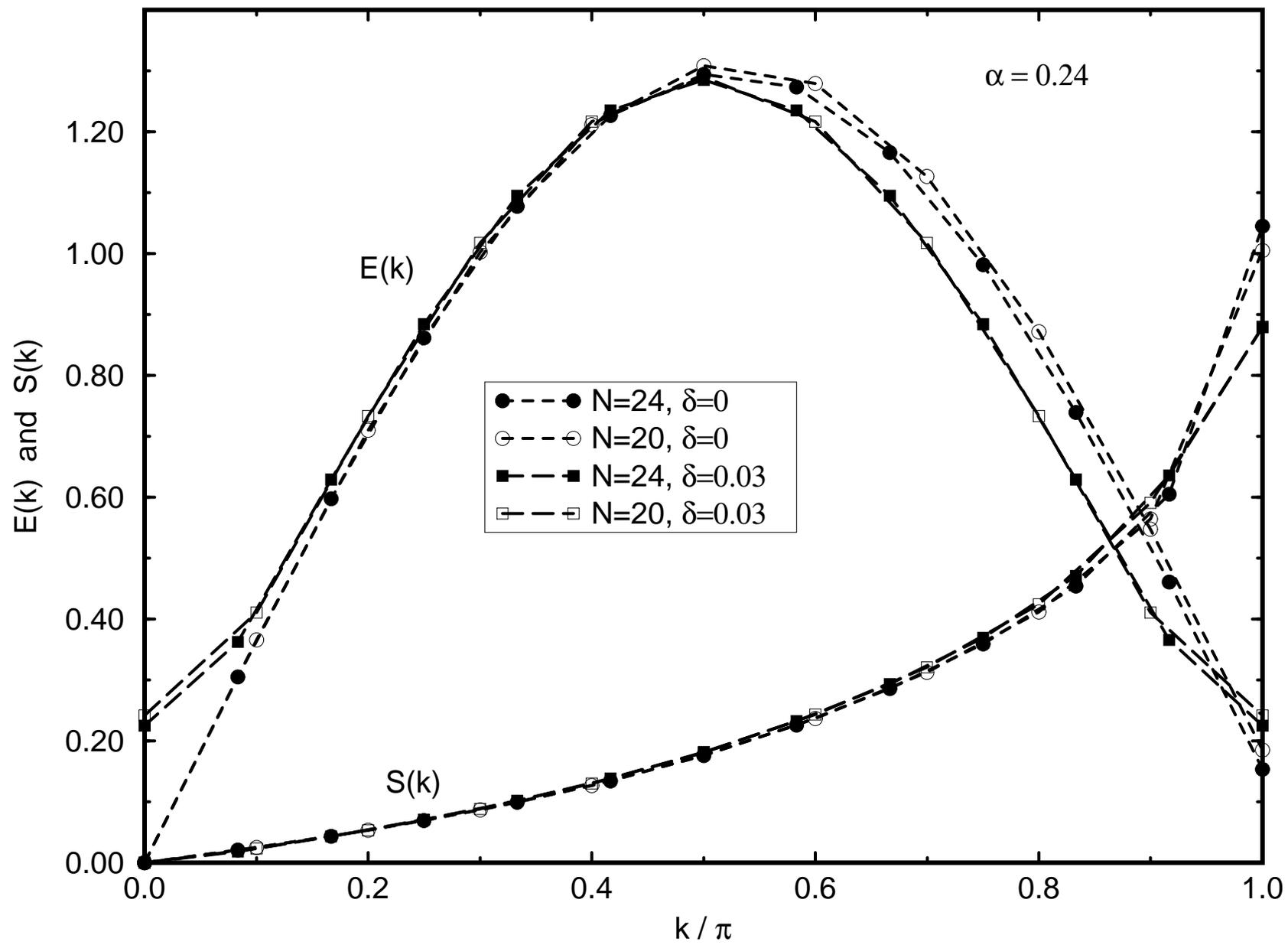
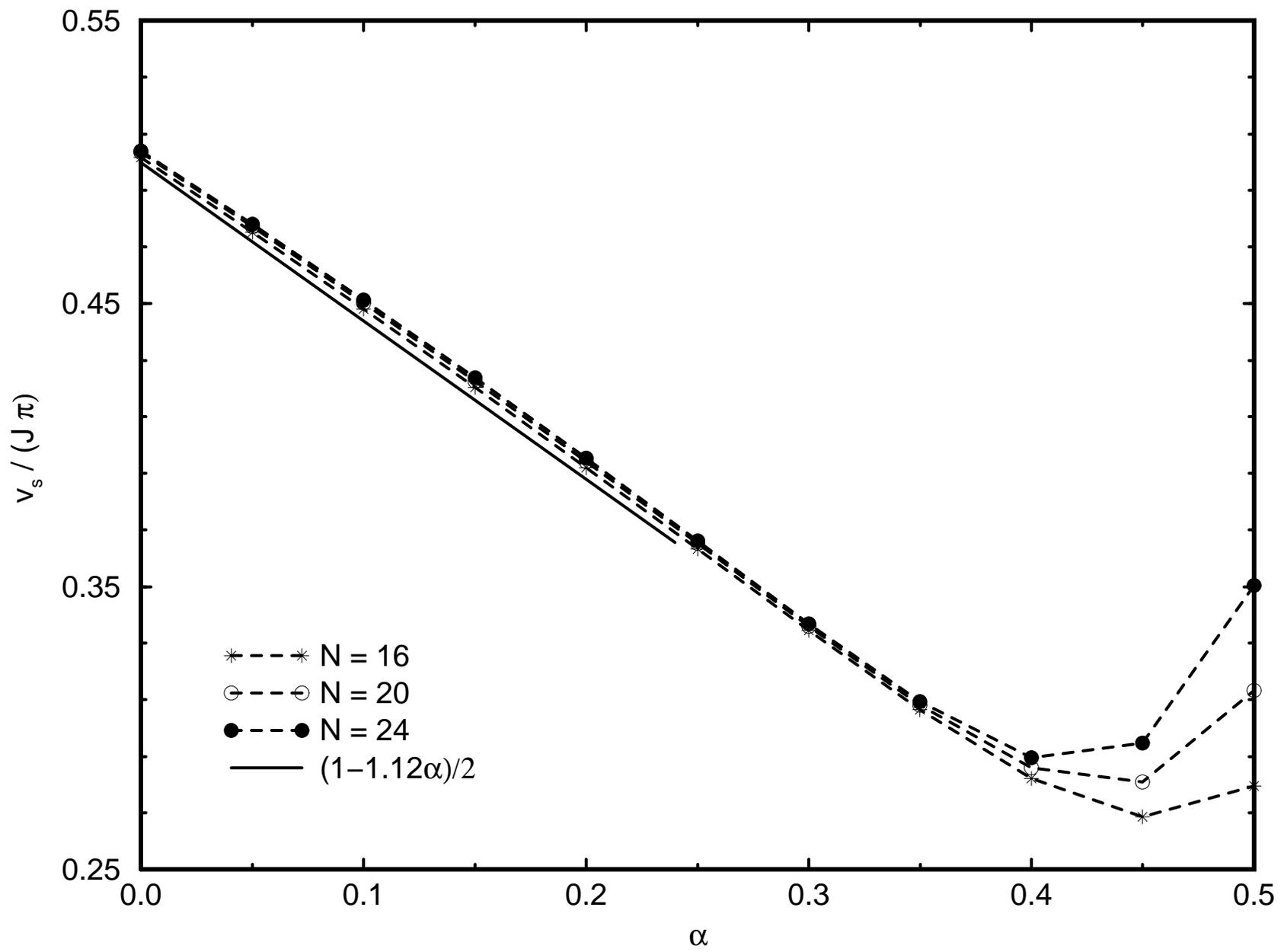
\end{document}